# BASIC_RV32s: An Open-Source Microarchitectural Roadmap for RISC-V RV32I

Hyun Woo Kang*[1] and Ji Woong Choi[2]
[1] Department of System Semiconductor Engineering, Sangmyung University, Cheonan, Republic of Korea
[2] Department of Computer Science and Engineering, Pohang University of Science and Technology, Pohang, Republic of Korea
*Email: hwctech1026@gmail.com

*Abstract*— This paper introduces BASIC_RV32s, an open-source framework providing a practical microarchitectural roadmap for the RISC-V RV32I architecture, addressing the gap between theoretical knowledge and hardware implementation. Following the classic Patterson and Hennessy methodology, the design evolves from a basic single-cycle core to a 5-stage pipelined core design with full hazard forwarding, dynamic branch prediction, and exception handling. For verification, the final core design is integrated into a System-on-Chip (SoC) with Universal Asynchronous Receiver-Transmitter (UART) communication implemented on a Xilinx Artix-7 Field-Programmable Gate Array (FPGA), achieving 1.09 Dhrystone million instructions per second per megahertz (DMIPS/MHz) at 50 MHz. By releasing all Register-Transfer Level (RTL) source code, signal-level logic block diagrams, and development logs under MIT license on GitHub, BASIC_RV32s offers a reproducible instructional pathway for the open-source hardware ecosystem.

*Keywords; RISC-V; Computer Architecture; FPGA; CPU*

## I. Introduction

The open-source RISC-V Instruction Set Architecture (ISA) offers significant flexibility for academic and commercial use, yet a practical gap remains between theoretical understanding and the implementation of functional processors. While many open-source cores exist, they often represent complex designs, making it difficult for learners to grasp the incremental development process. This paper presents BASIC_RV32s, a framework providing a systematic microarchitectural roadmap for RISC-V RV32I implementation. Following the Patterson and Hennessy methodology [1], our approach demonstrates step-by-step progression from a single-cycle core to a 5-stage pipelined processor verified on a dedicated System-on-Chip (SoC) implemented on Field-Programmable Gate Array (FPGA) hardware shown in Fig. 1. This work provides a documented instructional framework with complete Register-Transfer Level (RTL) sources and development history to facilitate processor design learning.

## II. From Single-Cycle to Pipelined Architecture

BASIC_RV32s follows three design principles to guide an intuitive and efficient hardware architecture design. 1) Streamline I/O signals to reduce complexity, 2) define clear module roles with focused logic to enhance modularity, and 3) prioritize performance-oriented features. The development began with RV32I37F as the foundational architecture. This single-cycle core implements 37 instructions from the RISC-V base integer set RV32I [2], excluding ECALL, EBREAK, and FENCE instructions to focus on core design. Features were added by introducing new modules, leading to RV32I43F, which adds Control and Status Register (CSR) access for the Zicsr extension [2], and RV32I46F, which adds a Trap Controller and Exception Detector to integrate the MRET instruction [3] and previously excluded ECALL, EBREAK to support trap mechanisms. The final design is RV32I46F_5SP, which transforms the single-cycle design into a 5-stage pipeline structure. The architecture of this pipelined core shown in Fig. 2 features a Hazard Unit to manage data and control hazards. Data hazards are resolved through forwarding from the execution (EX), memory access (MEM), and write-back (WB) stages [1]. Control hazards are handled by a 2-bit dynamic branch predictor in the Instruction Fetch stage, flushing the pipeline upon a misprediction in the EX stage.

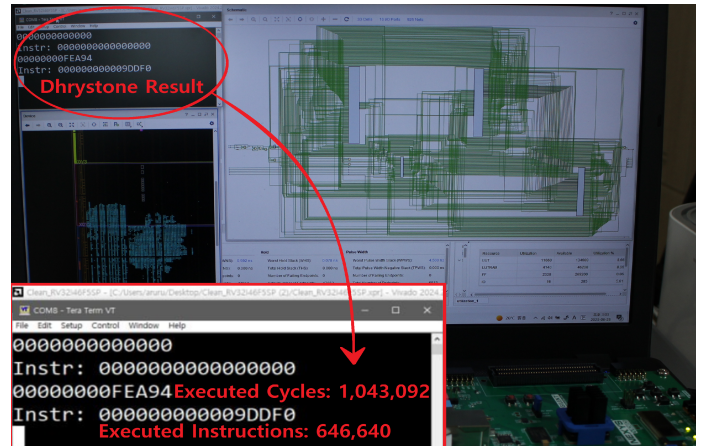

Fig. 1. RV32I46F_5SP core implemented on Nexys Video FPGA

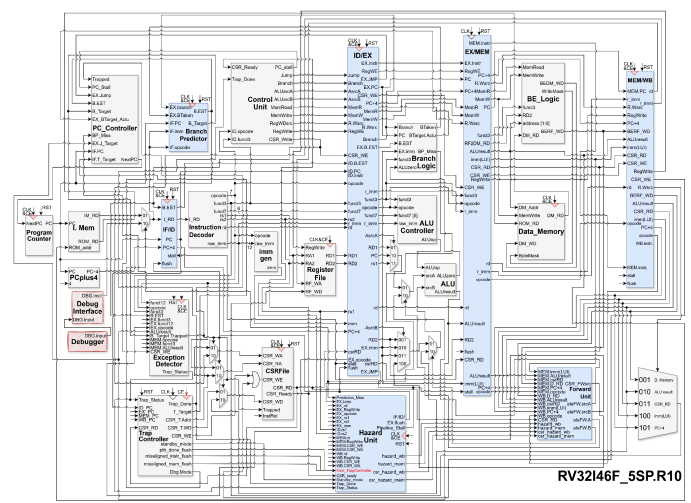

Fig. 2. Signal-level Block Diagram of RV32I46F_5SP microarchitecture

## III. SYSTEM-ON-CHIP DESIGN FOR CORE FPGA VERIFICATION

The 46F5SP_SoC architecture was designed for RV32I46F_5SP core FPGA verification. Fig. 3 shows the block diagram of the 46F5SP_SoC design. The SoC utilizes onboard General-Purpose Input/Output (GPIO) resources including 5 assigned buttons, a reset button, and 8 LEDs. The system uses Universal Asynchronous Receiver-Transmitter (UART) for debugging and benchmark output. Memory contains executable programs such as Dhrystone for performance evaluation. A Clock-enable signal was added to the core to support instruction-level debugging, allowing step-by-step execution controlled by button inputs. The 5 buttons provide benchmark control, debugging, and data monitoring of CPU operations. The system allows loading different programs into core memory while enabling customization of GPIO functions.

## IV. PERFORMANCE EVALUATION

The 46F5SP_SoC was synthesized with a 50 MHz timing constraint and verified at that frequency on a Digilent Nexys Video FPGA (Xilinx Artix-7 XC7A200T). Running Dhrystone 2.1 for 2,000 iterations with riscv-gcc -O2 and reading the mcycle, minstret CSRs, the core executed 646,640 instructions in 1,043,092 cycles (Fig. 1), corresponding to 1.09 Dhrystone million instructions per second per megahertz (DMIPS/MHz), and an average of 1.61 Cycles per Instruction. The standalone RV32I46F_5SP core occupies 3,010 Look-Up Tables (LUTs), and 998 Flip-Flops (FFs) as all memories are inferred as LUT-based distributed RAM. Adding GPIO and Dhrystone to form the 46F5SP_SoC expands its usage to 11,660 LUTs and 2,383 FFs. As detailed in Table I, RV32I46F_5SP delivers about 2.6 times higher DMIPS/MHz than NEORV32, likely attributed to 5-stage pipelining and dynamic branch prediction, representing competitive performance among educational RISC-V cores.

TABLE I. COMPARISON WITH OTHER EDUCATIONAL RISC-V CORES

| Processor | ISA | DMIPS /MHz | Frequency (MHz) | Pipeline Stages |
|---|---|---|---|---|
| RV32I46F_5SP (In-House) | RV32I Zicsr | 1.09 | 50 | 5 |
| NEORV32[a] (medium) | RV32IMC Zicsr Zifencei | 0.42 | 130 | 2 |
| DarkRISCV[b] | RV32I | 0.66 | 100 | 3 |

a. S. Nolting, "[Question] Dhrystone benchmark (Discussion #898)." GitHub Discussions (NEORV32). Accessed: Aug. 15, 2025. [Online]. Available: https://github.com/stnolting/neorv32/discussions/898
b. M. Samsoniuk, "DarkRISCV." GitHub, Accessed: Aug. 15, 2025. [Online]. Available: https://github.com/darklife/darkriscv

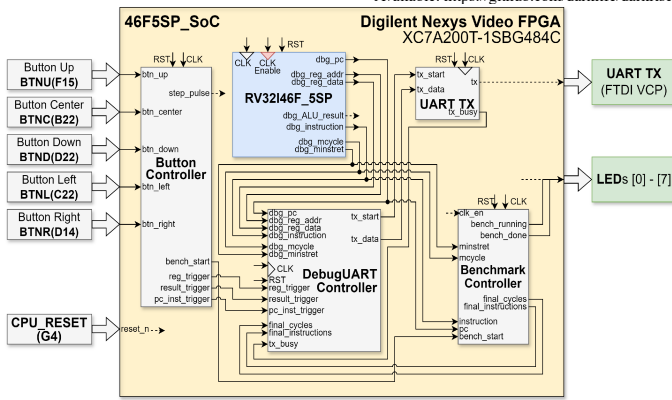

Fig. 3. Block Diagram of 46F5SP SoC Design

## V. GUIDELINES FOR MICROARCHITECTURE DESIGN AND IMPLEMENTATION CHALLENGES

BASIC_RV32s provides a toolkit that documents practical implementation challenges encountered during FPGA synthesis. Key challenges include combinatorial loop resolution through Finite State Machine redesign and asynchronous-to-synchronous module conversion, memory mapping conflicts solved by implementing bypass paths between instruction and data memory. The framework consists of four components: (1) incremental architecture documentation presenting each evolution with behavioral documentation of each logic module (2) dual-format Verilog code offering both clean and annotated versions with design rationale (3) development logs documenting debug traces, problem-solving approaches, and design iteration histories, and (4) verification methodologies providing UART-based debugging and Dhrystone benchmark setup on the FPGA. This systematic approach enables learners to design microarchitecture from scratch while understanding practical engineering challenges in hardware design. In recognition of its educational contribution, this work has been listed in the RISC-V official GitHub repository riscv/learn under Learning Resources as an Intermediate-Level Resource.

## VI. FUTURE WORK

Our future work includes implementing full ISA compliant exception handling according to the RISC-V Privileged Architecture [3], and hierarchical cache memory with RV64G extension for operating system support. We aim to extend this work into comprehensive guidelines for modern general-purpose processor design with advanced architectural features.

## VII. CONCLUSION

This paper presents BASIC_RV32s, an open-source instructional framework detailing the incremental design from a single-cycle to a pipelined RISC-V processor, and FPGA-verified SoC. The development logs, dual-format RTL source code, and signal-level block diagram of microarchitecture with behavioral documentation of each module demonstrate how theoretical concepts can be translated into working hardware. The resulting design and all associated materials are publicly available, offering a systematic instructional reference for the RISC-V community.

ACKNOWLEDGMENT

The authors thank Professor Y. W. Kim for guidance in RISC-V CPU design through his course and previous research. Special thanks to H. S. Kim and C. Kang for their support.